\documentclass[%
 reprint,
 amsmath,amssymb,
 aps,prl,
]{revtex4-2}

\usepackage{xcolor}

\usepackage{graphicx}
\usepackage{dcolumn}
\usepackage{bm}
\usepackage{hyperref}
\usepackage[normalem]{ulem}
\usepackage{multirow}

\begin{document}


\title{Studying 3D O($N$) Surface CFT on the Fuzzy Sphere}
\author{Jiechao Feng}
\affiliation{Graduate Group in Applied Science \& Technology, University of California, Berkeley, California 94720, USA \looseness=-2}

\author{Taige Wang}
\affiliation{Department of Physics, Harvard University, Cambridge, MA 02138, USA \looseness=-2}
\affiliation{Materials Research Laboratory, Massachusetts Institute of Technology, Cambridge, MA 02139, USA \looseness=-2}

\date{\today}

\begin{abstract}
Boundary conformal field theory (BCFT) provides a universal framework for critical phenomena in the presence of boundaries. 
We determine BCFT data for the normal and ordinary boundary universality classes of the $1+1$-dimensional boundaries of the $2+1$-dimensional $O(2)$ and $O(3)$ Wilson-Fisher fixed points, realized microscopically by a bilayer Heisenberg model on the fuzzy sphere. 
Using the fuzzy-sphere state-operator correspondence, we obtain boundary operator spectra, identify low-lying boundary primary operators, extract operator-product-expansion (OPE) data, and estimate the boundary central charges for both boundary conditions. 
For the normal boundary condition, the universal amplitudes $a_\sigma$ and $b_t$ extracted from one- and two-point functions agree quantitatively with Monte Carlo benchmarks where available. 
For both $N=2$ and $N=3$, we find a positive extraordinary-log exponent $\alpha$, providing independent microscopic evidence for extraordinary-log boundary criticality. 
Our results extend fuzzy-sphere BCFT spectroscopy beyond the Ising universality class to continuous $O(N)$ symmetry.
\end{abstract}

\maketitle

\textit{Introduction.}
Boundaries are common in physical systems and can change critical behavior.
In a conformal field theory (CFT), a boundary keeps only part of the bulk conformal symmetry.
The boundary can therefore flow to its own fixed point with universal data that are not fixed by the bulk alone.
Boundary critical phenomena are a classic subject~\cite{Diehl1997ThePhenomena,Pleimling04}, and have recently become a useful setting for conformal bootstrap, entanglement, and holography~\cite{Maldacena1999TheSupergravity,Qi11,Hasan10,Liendo2013TheCFTd,Gliozzi2015BoundaryBootstrap,Billo16,Lauria2018RadialCFTs,Mazac2019AnBCFTd,Behan20,Dey2020OperatorBCFT,Kaviraj2020TheCFT}.

The three-dimensional $O(N)$ model is one of the simplest examples.
For the ordinary surface transition the boundary preserves $O(N)$, while the extraordinary transition corresponds to boundary ordering at bulk criticality.
The special transition is the multicritical point between them.
Additionally, the normal transition is obtained by applying an explicit symmetry-breaking field at the boundary, which belongs to the same universality class as the extraordinary one when $d>3$~\cite{Bray77,Burkhardt87,Burkhardt94}.
For continuous symmetries in three dimensions, the extraordinary class is not obvious because a two-dimensional surface cannot have long-range $O(N\ge2)$ order by itself.
Recent field theory instead predicts an ``extraordinary-log'' surface universality class for $2\le N<N_c$~\cite{Metlitski2022BoundaryRevisited}, with surface order-parameter correlations that decay as $\langle \vec n(x)\cdot \vec n(0)\rangle\sim 1/(\log x)^q$.
Monte Carlo (MC) simulations support this picture for $N=2,3$~\cite{Parisen22,ParisenToldin2025UniversalModel,Hu2021Extraordinary-LogModel,ParisenToldin2021BoundaryClass}, while conformal bootstrap (CB) suggests $N_c\approx 5$~\cite{Padayasi2022TheBootstrap}.
The existence and exponent of the extraordinary-log phase can be determined from two universal amplitudes of the normal surface CFT~\cite{Metlitski2022BoundaryRevisited,Parisen22}.

Fuzzy-sphere regularization gives a direct way to study three-dimensional CFTs from a finite quantum many-body Hamiltonian~\cite{Madore1992TheSphere,Ippoliti2018Half-filledPoints,Zhu23}.
The method projects particles to the lowest Landau level on a sphere with a magnetic monopole and works at half filling.
At a critical point, the Hamiltonian realizes the CFT on $S^2\times\mathbb{R}$, where the state-operator correspondence maps energy eigenstates on the sphere to local CFT operators.
This approach has produced bulk CFT data~\cite{Hu2023OperatorSpheres,Voinea2025RegularizingSphere,Miro2025FlowingCFT,Zhou25a,Fan25,Zhou25b,He25,Yang25,Zhou24b,Han24,Zhou25c,Taylor2026ConformalSphere,Voinea2026CriticalTransition,Huffman26} and has also been applied to line defects~\cite{hu24,Zhou24,Cuomo24,Sarma2026FortuitousImpurities} and boundaries~\cite{Zhou25,Dedushenko24}, mainly in the three-dimensional Ising universality class.

In this Letter, we use the fuzzy-sphere framework to study the normal and ordinary surface CFTs of the three-dimensional $O(N)$ model for $N=2,3$.
The main difference from classical MC is that we do not simulate a three-dimensional slab.
Instead, we diagonalize a quantum Hamiltonian on the sphere and read boundary operator multiplets directly from the spectra.
This gives access not only to the amplitudes $a_\sigma$ and $b_t$, but also to boundary descendants, previously unknown primaries, and the boundary central charge in the same microscopic calculation.
Our normal-boundary data agree with available MC benchmarks and give $\alpha>0$ for both $N=2$ and $N=3$, supporting the extraordinary-log scenario from an independent regularization.

\textit{Model.}
We start from the fuzzy-sphere Hamiltonian of Ref.~\cite{Han24}, which realizes the bulk $O(3)$ CFT in a bilayer Heisenberg magnet,
\begin{align}
\label{eq:H_O3}
    H_{\rm int} &= \int d\boldsymbol{\Omega}_{a,b}\, \Big\{ U_0 n(\boldsymbol{\Omega}_a)n(\boldsymbol{\Omega}_b)
    + U_2 \boldsymbol{n}_1(\boldsymbol{\Omega}_a)\cdot \boldsymbol{n}_2(\boldsymbol{\Omega}_b) \nonumber \\
    &\quad -U_1\big[\boldsymbol{n}_1(\boldsymbol{\Omega}_a)\cdot\boldsymbol{n}_1(\boldsymbol{\Omega}_b)
    +\boldsymbol{n}_2(\boldsymbol{\Omega}_a)\cdot\boldsymbol{n}_2(\boldsymbol{\Omega}_b)\big]\Big\} \nonumber \\
    &\quad - h\int d\boldsymbol{\Omega}\, \hat{\boldsymbol{\Psi}}^{\dagger}\tau^x\sigma^0\hat{\boldsymbol{\Psi}} .
\end{align}
Here $\boldsymbol{\Psi}=(\psi_{1\uparrow},\psi_{1\downarrow},\psi_{2\uparrow},\psi_{2\downarrow})$ is a four-flavor fermion, $\tau$ labels the layer, and $\sigma$ labels spin.
The spin density on layer $\tau$ is $\boldsymbol{n}_\tau(\boldsymbol{\Omega})=\psi_\tau^\dagger(\boldsymbol{\Omega})\boldsymbol{\sigma}\psi_\tau(\boldsymbol{\Omega})$, while $n(\boldsymbol{\Omega})=\boldsymbol{\Psi}^\dagger(\boldsymbol{\Omega})\boldsymbol{\Psi}(\boldsymbol{\Omega})$ is the total density.
To obtain the $O(2)$ bulk CFT, we add anisotropic intra- and interlayer interactions that break the global $O(3)$ symmetry down to $O(2)$ and tune to the corresponding critical point.
As a check for a genuine bulk critical point, Appendix~\ref{app:OPE_O2} gives four operator product expansion (OPE) coefficients in $O(2)$ CFT from the fuzzy sphere.
They agree well with conformal bootstrap values and have not previously been reported from this regularization.

Surface criticality is implemented by modifying orbitals with magnetic quantum number $m<0$, which realizes an orbital-space boundary~\cite{Zhou25}.
We have checked that this construction flows to the same surface universality class as a real-space pinning field on the southern hemisphere (See Appendix~\ref{app:real_space_cut}).
For the normal surface CFT we add the one-body term
\begin{equation}
\label{eq:surface_orbital}
    H_{\rm surface}=\lim_{h\to\infty}\sum_{m<0} h\, \mathbf{c}_m^\dagger \tau^z\sigma^z\mathbf{c}_m ,
\end{equation}
where $\mathbf{c}_m$ is the four-flavor fermion and $m=-s,-s+1,\ldots,s$ is the orbital index.
For the $O(2)$ normal surface CFT, the pinning direction is chosen within the two-component $O(2)$ order parameter.
For the ordinary surface CFT, we instead empty the $m<0$ orbitals, preserving the global $O(N)$ symmetry.
In both cases, the remaining $m>0$ orbitals are kept at half filling.
All numerical results were obtained using FuzzifiED~\cite{Zhou2025FuzzifiED:Sphere}.

\begin{figure}[t]
\includegraphics[width=\columnwidth]{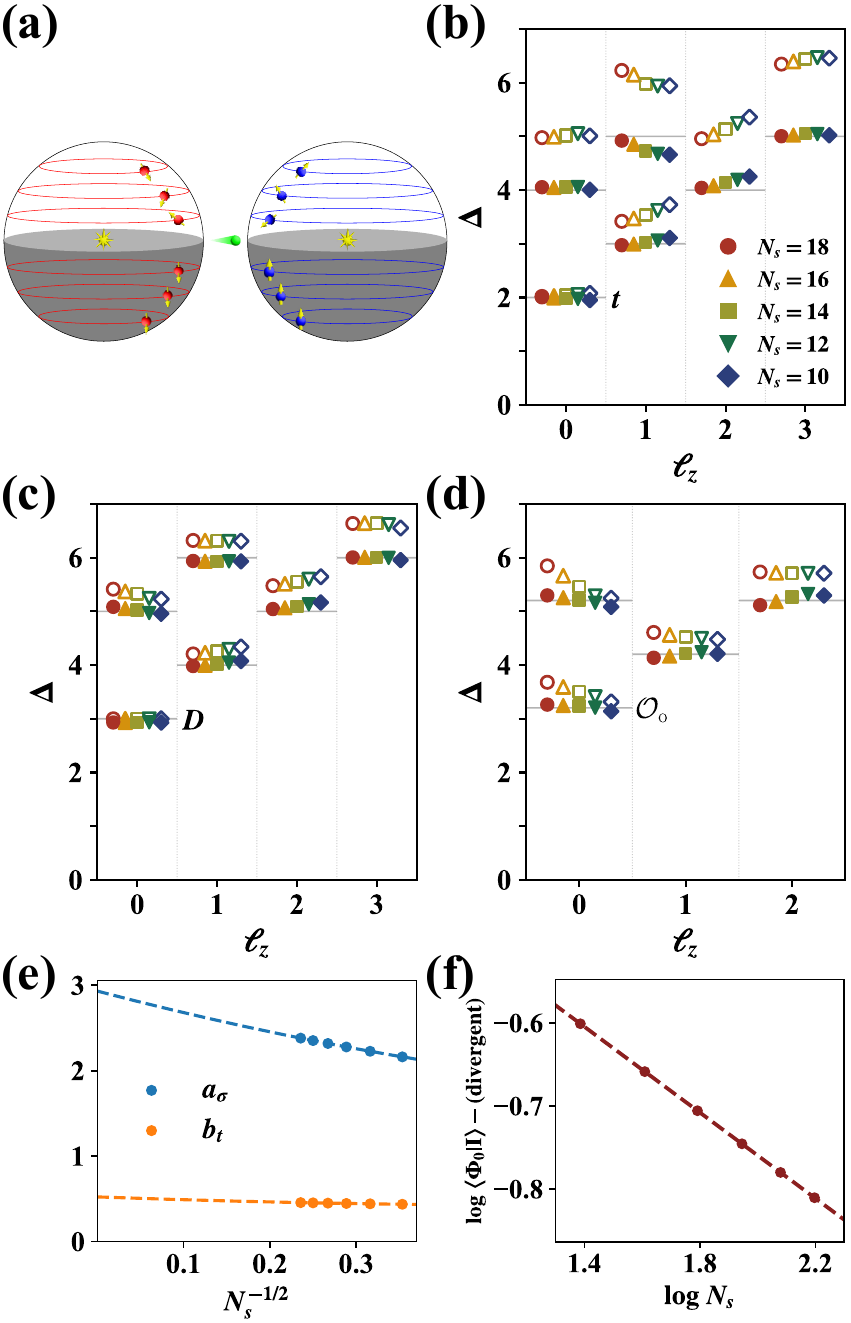}
\caption{(a) Schematic realization of an $O(N)$ normal surface CFT on the fuzzy sphere. (b)-(f) CFT data for the $O(2)$ normal surface CFT. Conformal multiplets are shown for the surface primaries (b) $t$, (c) $D$, and (d) $\mathcal{O}_{\rm o}$ for $10\le N_s\le 18$. Open symbols are raw dimensions calibrated by $\Delta_D=3$. Solid symbols include the conformal-perturbation-theory correction from the leading irrelevant boundary perturbation associated with $D$. (e) Finite-size extrapolation of the universal amplitudes $a_\sigma$ and $b_t$. (f) Finite-size scaling of the overlap between the bulk CFT ground state and the polarized state $|\Phi_0\rangle$, from which $c_{\rm nor}$ is extracted.}
\label{fig:Fig1}
\end{figure}

\textit{Normal surface CFTs.}
A planar boundary breaks the bulk conformal symmetry from $SO(4,1)$ to $SO(3,1)$.
Boundary local operators therefore form primaries and descendants of the two-dimensional boundary conformal algebra.
For a boundary primary $\hat{\phi}$ of dimension $\Delta_{\hat\phi}$, descendants have the form $\partial^m\bar\partial^n\hat\phi$, scaling dimensions $\Delta_{\hat\phi}+m+n$, and angular momenta $l_z=m-n$.
In finite systems we identify the conformal multiplet for displacement operator $D$ and use its protected value $\Delta_D=3$ to set the nonuniversal speed of light~\cite{Zhou25}~\footnote{We have verified that using the conserved bulk stress tensor to calibrate the speed of light yields consistent results, supporting the claim that the boundary speed of light is identical to that in the bulk. Details are given in Appendix~\ref{app:calibration}.}.
This converts energy gaps into scaling dimensions which are reported as raw scaling dimensions below.

For a normal $O(N)$ surface CFT, the boundary field selects a pinning axis and breaks the global symmetry from $O(N)$ to $O(N-1)$.
For $N=3$, boundary operators can therefore be further labeled by the residual $O(2)$ charge $S$ about this axis.
For $N=2$, they can instead be labeled by the $O(1)\cong\mathbb{Z}_2$ charge. 
Among the low-lying primaries, the protected operators are the tilt operator $t$, an $O(N-1)$ vector with $\Delta_t=2$, and the displacement operator $D$, an $O(N-1)$ scalar with $\Delta_D=3$~\cite{Cardy1990UniversalGeometries,Bray77,Burkhardt87,Parisen22,Padayasi2022TheBootstrap}.
Their dimensions are fixed by Ward identities associated with the corresponding broken generators: $t$ is tied to the broken $O(N)/O(N-1)$ rotations, while $D$ appears in the displacement normal to the boundary.

\begin{table}[t]
\caption{CFT data for the $O(2)$ and $O(3)$ normal surface CFTs.}
\label{tab:normal}
\begin{ruledtabular}
\renewcommand{\arraystretch}{1.18}
\begin{tabular}{clcccc}
 & & This work & MC~\cite{Parisen22} & CB~\cite{Padayasi2022TheBootstrap} & $\epsilon$-exp.~\cite{Giombi2025HigherCFT} \\
\hline
$N=2$ & $\Delta_t$      & 2.00(1)  &        & 2 &        \\
      & $\Delta_D$      & 2.932(2) &        & 3 &        \\
      & $\Delta_{\mathcal{O}_{\rm o}}$  & 3.26(2)  &        &   &        \\
      & $\Delta_{\mathcal{O}_{\rm e}}$  & 3.58(5)  &        &   &        \\
\cline{2-6}
      & $a_{\sigma}$    & 2.932(1) & 2.880(2) & 2.923 & 3.01169 \\
      & $b_t$           & 0.523(2) & 0.525(4) & 0.4882 &        \\
      & $\alpha$        & 0.313(2) & 0.300(5) & 0.3567 &        \\
      & $C_t$           & 0.199(2) & 0.191(3) & 0.2271 &        \\
\cline{2-6}
      & $c_{\rm nor}$   & $-1.550(3)$ &        &        & $-1.63$ \\
\hline
$N=3$ & $\Delta_t$          & 2.05(1)  &        & 2 &        \\
      & $\Delta_D$          & 2.877(3) &        & 3 &        \\
      & $\Delta_{\mathcal{O}_{S=0}}$  & 3.52(12) &        &   &        \\
      & $\Delta_{\mathcal{O}_{S=1}}$  & 3.56(5)  &        &   &        \\
      & $\Delta_{\mathcal{O}_{S=2}}$  & 3.33(11) &        &   &        \\
\cline{2-6}
      & $a_{\sigma}$    & 3.147(5) & 3.136(2) & 3.16 & 3.22339 \\
      & $b_t$           & 0.533(6) & 0.529(3) & 0.5092 &        \\
      & $\alpha$        & 0.188(8) & 0.190(4) & 0.2237 &        \\
      & $C_t$           & 0.221(5) & 0.223(3) & 0.2437 &        \\
\cline{2-6}
      & $c_{\rm nor}$   & $-1.913(5)$ &        &        & $-1.92$ \\
\end{tabular}
\end{ruledtabular}
\end{table}

We first discuss the $O(2)$ normal surface CFT.
Fig.~\ref{fig:Fig1}(b)-(d) and Fig.~\ref{fig:5}(a) show its lowest four conformal multiplets up to $\Delta=6$ and $l_z=3$ for $10\le N_s\le18$.
Open symbols denote raw spectra normalized by $\Delta_D=3$ at each $N_s$.
Solid symbols include the leading correction from the irrelevant boundary perturbation associated with $D$, computed using conformal perturbation theory as detailed in Appendix~\ref{app:CPT}~\cite{Lao20233DTheory}.
After this correction the descendants show the expected integer spacing.
Beyond the protected operators $t$ and $D$, we identify two additional primaries, denoted $\mathcal{O}_{\rm o}$ and $\mathcal{O}_{\rm e}$, with
\begin{equation}
    \Delta_{\mathcal{O}_{\rm o}}=3.26(2),\qquad \Delta_{\mathcal{O}_{\rm e}}=3.58(5).
\end{equation}
The quoted uncertainties are estimated from the difference between the corrected values at $N_s=18$ and $N_s=16$.
$\mathcal{O}_{\rm o}$ is the lowest $\mathbb{Z}_2$-odd primary above $t$, while $\mathcal{O}_{\rm e}$ is the lowest $\mathbb{Z}_2$-even primary above $D$,

We next extract OPE data for the same $O(2)$ normal surface CFT.
Writing the bulk order parameter as $\vec\phi=(\vec\varphi,\sigma)$, with $\sigma$ parallel to the boundary field, the one-point function of $\sigma$ and the bulk-to-boundary OPE into the tilt operator define
\begin{equation}
    \langle \sigma(\mathbf{r})\rangle = \frac{a_\sigma}{(2z)^{\Delta_\phi}}, \qquad
    \langle t^i(0)\varphi^j(\mathbf{r})\rangle
    =\delta_{ij}b_t\frac{(2z)^{2-\Delta_\phi}}{|\mathbf{r}|^4},
\end{equation}
where $\Delta_\phi$ is the scaling dimension of the leading bulk $O(N)$ vector~\cite{Metlitski2022BoundaryRevisited,Parisen22}.
The combination
\begin{equation}
    \alpha \equiv \frac{\pi}{2}\left(\frac{a_\sigma}{4\pi b_t}\right)^2-\frac{N-2}{2\pi}
\label{eq:alpha}
\end{equation}
controls the extraordinary-log phase, which exists for $\alpha>0$~\cite{Metlitski2022BoundaryRevisited,Parisen22}.
Its logarithmic exponent is $q=(N-1)/(2\pi\alpha)$.
The same ratio defines the current-tilt coefficient
\begin{equation}
    C_t^{1/2}=\frac{a_\sigma}{4\pi b_t},
\label{eq:c_t}
\end{equation}
through the boundary limit of the broken $O(N)$ current, $j^z_{[Ni]}(\boldsymbol{r})\sim \sqrt{C_t}\,t_i(\boldsymbol{r})$ as $z\to0$, with $i=1,\ldots,N-1$~\cite{Padayasi2022TheBootstrap}.

For $N=2$, Fig.~\ref{fig:Fig1}(e) shows finite-size extrapolations using the scaling form of Ref.~\cite{Zhou25},
\begin{subequations}
\label{eq:universal}
\begin{align}
    a_\sigma &: \quad 2.932(1)-2.652N_s^{-1/2}+1.344N_s^{-1}, \\
    b_t &: \quad 0.523(2)-0.363N_s^{-1/2}+0.338N_s^{-1}.
\end{align}
\end{subequations}
The uncertainties are estimated by comparing fits with and without the largest included system size.
They should therefore be viewed as finite-size fitting errors rather than full systematic uncertainties.
The extrapolated values agree with MC within about $2\%$, $a_\sigma=2.880(2)$ and $b_t=0.525(4)$~\cite{Parisen22}.
Eqs.~\eqref{eq:alpha} and~\eqref{eq:c_t} then give $\alpha=0.313(2)$ and $C_t=0.199(2)$.
The positive value of $\alpha$ supports the extraordinary-log phase for $N=2$.
It is consistent with $\alpha=0.300(5)$ from normal-boundary MC~\cite{Parisen22} and $\alpha_{\rm eo}=0.27(2)$ from direct simulations in the extraordinary region~\cite{Hu2021Extraordinary-LogModel}.

\begin{figure}[t]
\includegraphics[width=\columnwidth]{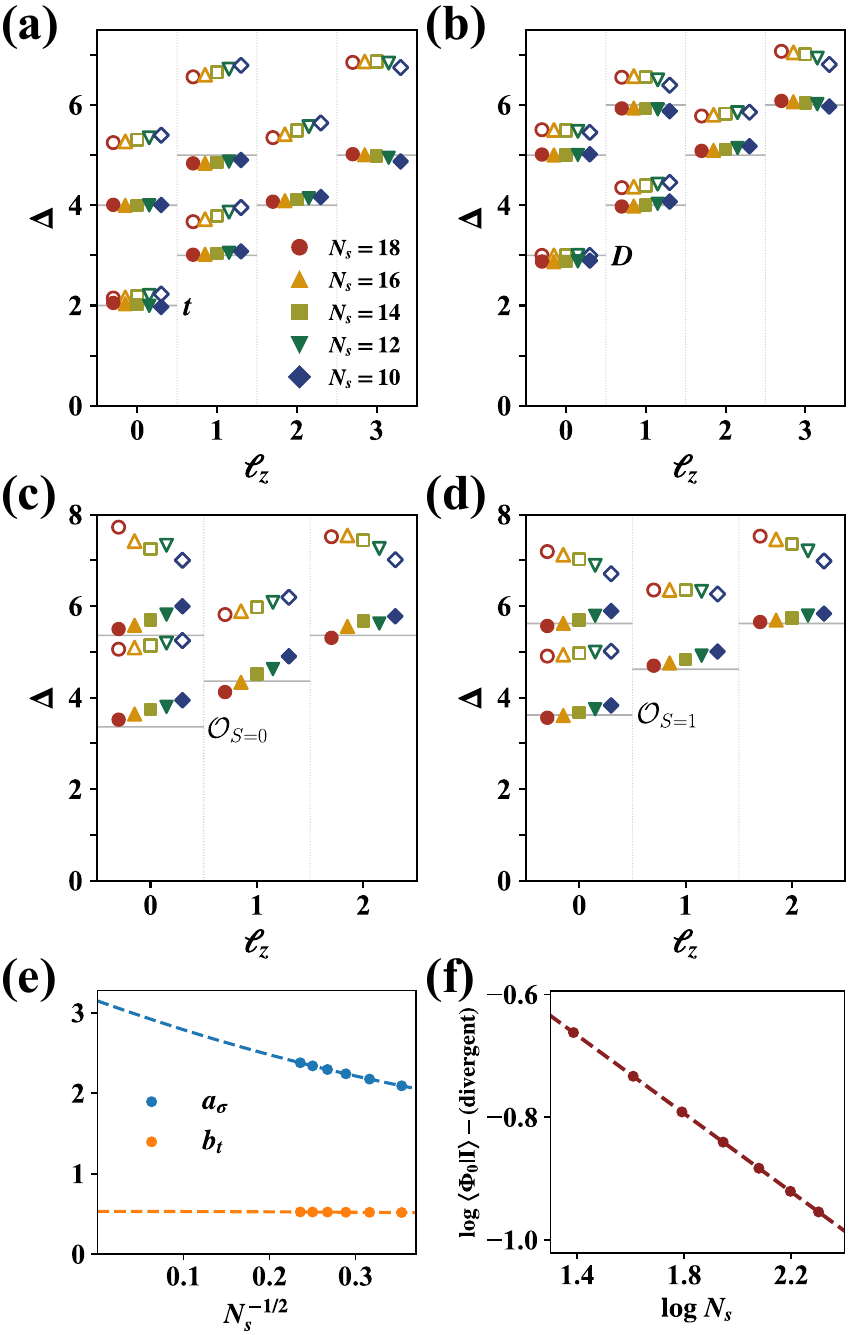}
\caption{CFT data for the $O(3)$ normal surface CFT. Conformal multiplets are shown for (a) $t$, (b) $D$, (c) $\mathcal{O}_{S=0}$, and (d) $\mathcal{O}_{S=1}$ for $10\le N_s\le18$. (e) Finite-size extrapolation of the universal amplitudes $a_\sigma$ and $b_t$. (f) Finite-size scaling of the overlap between the bulk CFT ground state and the polarized state $|\Phi_0\rangle$, yielding $c_{\rm nor}$.}
\label{fig:Fig2}
\end{figure}

Fig.~\ref{fig:Fig1}(f) shows the finite-size scaling of the wavefunction overlap between the bulk ground state $|\mathbb{I}\rangle$ and the polarized state $|\Phi_0\rangle$ that implements the normal boundary condition.
Following Ref.~\cite{Zhou25},
\begin{equation}
    \log\langle \mathbb{I}|\Phi_0\rangle
    = C_1N+C_0+C_{-1}/N+(c_{\rm bd}/6)\log N,
\end{equation}
where the $C_i$ are nonuniversal constants and $c_{\rm bd}$ is the universal boundary central charge.
In three-dimensional surface CFT, $c_{\rm bd}$ is the coefficient of a boundary conformal anomaly, or equivalently the logarithmic term in the hemisphere free energy~\cite{Jensen2016ConstraintFlows,Herzog18,Giombi2025HigherCFT}.
It is not the Virasoro central charge of a standalone two-dimensional CFT, so a negative value is not a sign of nonunitarity.
For example, with the convention of Ref.~\cite{Giombi2025HigherCFT}, a free scalar with Dirichlet boundary condition has $c_{\rm bd}<0$.
Fitting data from $N_s=4$ to $9$ yields $c_{\rm nor}=-1.550(3)$, close to the two-loop AdS $\epsilon$-expansion estimate $c_{\rm nor}\simeq -1.63$~\cite{Giombi2025HigherCFT}.

We now repeat the same analysis for the $O(3)$ normal surface CFT.
Fig.~\ref{fig:Fig2}(a)-(d) and Fig.~\ref{fig:5}(b) show conformal multiplets of $t$, $D$, $\mathcal{O}_{S=0}$, $\mathcal{O}_{S=1}$, and $\mathcal{O}_{S=2}$ up to $\Delta=6$ and $l_z=3$.
After removing the leading finite-size correction associated with $D$, the corrected spectra are again consistent with integer-spaced descendants.
In addition to $t$ and $D$, we find three previously unidentified primaries: the lowest primary above $D$ in the $S=0$ sector, $\mathcal{O}_{S=0}$, with $\Delta_{\mathcal{O}_{S=0}}=3.52(12)$; the lowest primary above $t$ in the $S=1$ sector, $\mathcal{O}_{S=1}$, with $\Delta_{\mathcal{O}_{S=1}}=3.56(5)$; and the lowest primary in the $S=2$ sector, $\mathcal{O}_{S=2}$, with $\Delta_{\mathcal{O}_{S=2}}=3.33(11)$.
Together with the $O(2)$ results, this supports the conjecture that the normal surface spectrum contains no operator lighter than the displacement operator except for the tilt~\cite{Padayasi2022TheBootstrap,Bray77,Dey2020OperatorBCFT}.
We also checked that, for $N=3$, the lowest primaries in sectors with $S>2$ all have dimensions larger than $3$.

For the $O(3)$ OPE data, Fig.~\ref{fig:Fig2}(e) gives
\begin{subequations}
\begin{align}
    a_\sigma &: \quad 3.147(5)-3.815N_s^{-1/2}+2.356N_s^{-1}, \\
    b_t &: \quad 0.533(6)-0.014N_s^{-1/2}-0.078N_s^{-1}.
\end{align}
\end{subequations}
These extrapolated values agree with MC within about $1\%$, $a_\sigma=3.136(2)$ and $b_t=0.529(3)$~\cite{Parisen22}.
They imply $\alpha=0.188(8)$ and $C_t=0.221(5)$, in good agreement with $\alpha=0.190(4)$ from normal-boundary MC~\cite{Parisen22} and compatible with $\alpha_{\rm eo}=0.15(2)$ from direct extraordinary-log simulations~\cite{ParisenToldin2021BoundaryClass}.
Thus $\alpha>0$ again supports the extraordinary-log phase, now for $N=3$.
Finally, the overlap scaling in Fig.~\ref{fig:Fig2}(f), using data from $N_s=4$ to $10$, gives $c_{\rm nor}=-1.913(5)$, consistent with the $\epsilon$-expansion estimate $c_{\rm nor}\simeq -1.92$~\cite{Giombi2025HigherCFT}.

\textit{Ordinary surface CFTs.}
We now turn to the ordinary boundary condition, which preserves the global $O(N)$ symmetry.
Boundary operators are therefore labeled by the same $O(N)$ quantum numbers (denoted as $S$) as bulk operators.
Additionally, for the $S=0$ sector of $N=2$ and all $S$ sectors of $N=3$, there is a $\mathbb{Z}_2$ symmetry which swaps layer 1 and 2~\cite{Guo2025TheSphere}.
The lowest $O(N)$-singlet, $\mathbb{Z}_2$-even primary is the displacement operator $D$ with protected dimension $3$~\cite{Giombi2020CFTFlows}.
The leading $O(N)$-vector boundary field $\hat\phi$ is the only relevant surface operator~\cite{Deng2005SurfaceModels,ParisenToldin2023TheModel}.
Unlike $D$, its dimension is not protected.

\begin{figure}[t]
\includegraphics[width=\columnwidth]{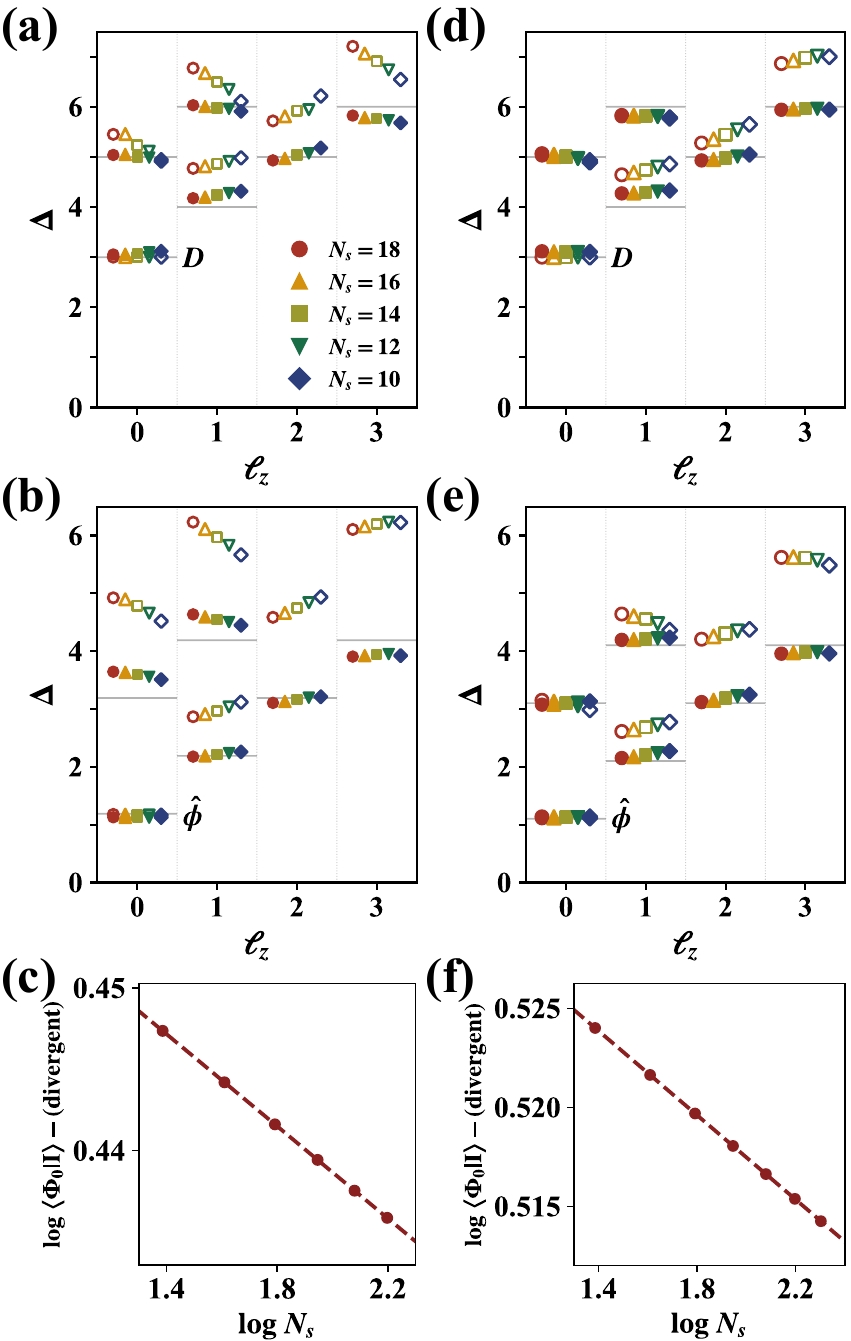}
\caption{CFT data for the $O(2)$ and $O(3)$ ordinary surface CFTs. Conformal multiplets are shown for (a) $D$ and (b) $\hat\phi$ in the $O(2)$ ordinary surface CFT, and for (d) $D$ and (e) $\hat\phi$ in the $O(3)$ ordinary surface CFT. Panels (c) and (f) show finite-size scaling of the overlap between the bulk CFT ground state and the polarized state $|\Phi_0\rangle$, from which $c_{\rm ord}$ is obtained.}
\label{fig:Fig3}
\end{figure}

\begin{table}[t]
\caption{CFT data for the $O(2)$ and $O(3)$ ordinary surface CFTs.}
\label{tab:ordinary}
\begin{ruledtabular}
\renewcommand{\arraystretch}{1.18}
\begin{tabular}{clcccc}
 & & This work & MC~\cite{ParisenToldin2023TheModel} & CB~\cite{Gliozzi2015BoundaryBootstrap} & $\epsilon$-exp.~\cite{Diehl1986Field-theoreticSurfaces,Giombi2025HigherCFT} \\
\hline
$N=2$ & $\Delta_{\hat\phi}$ & 1.128(3) & 1.2286(25) & 1.2342(9) & 1.19 \\
      & $\Delta_{\mathcal{O}}$          & 2.83(7)  &            &           &      \\
      & $\Delta_{D}$        & 3.053(6) &            &           &      \\
\cline{2-6}
      & $c_{\rm ord}$       & $-0.0851(5)$ &       &           & $-0.055$ \\
\hline
$N=3$ & $\Delta_{\hat\phi}$ & 1.112(3) & 1.194(3) & 1.198(1) & 1.153 \\
      & $\Delta_{\mathcal{O}}$          & 2.499(8) &          &          &       \\
      & $\Delta_{D}$        & 3.117(8) &          &          &       \\
\cline{2-6}
      & $c_{\rm ord}$       & $-0.064(3)$ &       &          & $-0.072$ \\
\end{tabular}
\end{ruledtabular}
\end{table}

Fig.~\ref{fig:Fig3}(a) and Fig.~\ref{fig:Fig3}(d) show the displacement multiplets of the $O(2)$ and $O(3)$ ordinary surface CFTs up to $\Delta=6$ and $l_z=3$.
Fig.~\ref{fig:Fig3}(b) and Fig.~\ref{fig:Fig3}(e) show the corresponding multiplets of $\hat\phi$, from which we obtain
\begin{equation}
    \Delta_{\hat\phi}=1.128(3)\quad (N=2),\qquad
    \Delta_{\hat\phi}=1.112(3)\quad (N=3).
\end{equation}
For comparison, high-precision MC gives $\Delta_{\hat\phi}=1.2286(25)$ for $N=2$ and $1.194(3)$ for $N=3$~\cite{ParisenToldin2023TheModel}, while conformal bootstrap gives $1.2342(9)$ and $1.198(1)$, respectively~\cite{Gliozzi2015BoundaryBootstrap}.
We also identify the lowest $\mathbb{Z}_2$-odd $S=0$ primary in the $O(2)$ ordinary surface CFT, denoted $\mathcal{O}$, with $\Delta_{\mathcal{O}}=2.83(7)$ [Fig.~\ref{fig:5}(c)].
For the $O(3)$ ordinary surface CFT, the lowest $\mathbb{Z}_2$-even $S=2$ primary has $\Delta_{\mathcal{O}}=2.499(8)$ [Fig.~\ref{fig:5}(d)].

The boundary central charge for the ordinary class is obtained from the same overlap scaling scheme as in the normal case.
Here we use a symmetry-preserving polarized state in which one layer is fully occupied by both spin flavors and the other layer is empty.
Fig.~\ref{fig:Fig3}(c) and Fig.~\ref{fig:Fig3}(f) give
\begin{equation}
\begin{gathered}
    c_{\rm ord}=-0.0851(5)\quad (N=2),\\
    c_{\rm ord}=-0.064(3)\quad (N=3).
\end{gathered}
\end{equation}
These values may be compared with the $\epsilon$-expansion estimates $c_{\rm ord}\simeq -0.055$ for $N=2$ and $c_{\rm ord}\simeq -0.072$ for $N=3$~\cite{Giombi2025HigherCFT}.

\textit{Discussion and summary.}
Tables~\ref{tab:normal} and~\ref{tab:ordinary} summarize the main CFT data obtained in this work.
For the normal surface CFTs, both the operator spectra and the OPE data are consistent with the expected protected structure and with available MC benchmarks.
In particular, the fuzzy-sphere estimates of $a_\sigma$ and $b_t$ imply $\alpha>0$ for $N=2,3$.
This gives an independent microscopic check of the extraordinary-log scenario.
The seven newly identified surface primaries all have $\Delta>2$, and are therefore irrelevant on the two-dimensional boundary.

The ordinary surface CFTs show larger finite-size corrections, especially for the $O(2)$ model.
This trend is consistent with the quality of the underlying bulk critical Hamiltonians.
The $O(3)$ pseudopotentials appear to be more finely tuned, which reduces corrections from irrelevant bulk and boundary perturbations compared with the present $O(2)$ implementation.
Further optimization of the $O(2)$ bulk Hamiltonian and larger-system calculations, for example using DMRG, should improve the ordinary-boundary estimates and allow access to higher boundary operators and additional OPE coefficients.

More generally, this work extends fuzzy-sphere boundary methods from the Ising case to continuous $O(N)$ symmetry.
Applying the same normal-boundary analysis to $N=4$ and $N=5$ would be especially useful, because it would probe the critical value $N_c$ that marks the end of the extraordinary-log phase~\cite{Guo2025TheSphere,Dey2025ConformalModel}.

\textit{Note added.} When preparing this manuscript, we became aware of another work reporting OPE coefficients in O(2) bulk CFT using the fuzzy sphere method~\cite{Dey26}. Among the four OPE values reported here, two were also reported in their work: $f_{\phi \phi s}$ and $f_{sss}$.

\textbf{Acknowledgments.}
We thank Wei Zhu, Haoran Cui, Michael Zaletel, and Max Metlitski for helpful discussions. J.F. was funded by the U.S. Department of Energy, Office of Science, Office of Basic Energy Sciences, Materials Sciences and Engineering Division under Contract No. DE-AC02-05-CH11231 (Theory of Materials program KC2301). T.W. is grateful for the support by the Harvard Quantum Initiative Fellowship and the Simons Collaboration on Ultra-Quantum Matter, which is a grant from the Simons Foundation (Grant No. 651440). This research uses the Lawrencium computational cluster provided by the Lawrence Berkeley National Laboratory (supported by the U.S. Department of Energy, Office of Basic Energy Sciences under Contract No. DE-AC02-05-CH11231).

\bibliography{reference}

\newpage
\setcounter{secnumdepth}{3}
\appendix

\section{Bulk $O(2)$ OPE coefficients}
\label{app:OPE_O2}

The bilayer Heisenberg model used for the $O(3)$ Wilson-Fisher CFT can be tuned to the $O(2)$ Wilson-Fisher CFT by adding an anisotropy to the spin interactions~\cite{Han24,Yang25}.
We determine the critical parameters by minimizing a cost function that measures the deviation of the finite-size spectrum from reference CFT data~\cite{Guo2025TheSphere}.
Relative to Eq.~\eqref{eq:H_O3}, the anisotropy is added to the $y$ component in both the intralayer and interlayer interactions.
The Hamiltonian used for the bulk $O(2)$ CFT is
\begin{align}
\label{eq:H_O2_app}
    H_{\rm int}^{O(2)} &= \int d\boldsymbol{\Omega}_{a}d\boldsymbol{\Omega}_{b}
    \bigg\{ U_0 n(\boldsymbol{\Omega}_a)n(\boldsymbol{\Omega}_b) \nonumber \\
    &\quad + U_2 \big[n_{1x}(\boldsymbol{\Omega}_a)n_{2x}(\boldsymbol{\Omega}_b)
    + n_{1z}(\boldsymbol{\Omega}_a)n_{2z}(\boldsymbol{\Omega}_b) \nonumber \\
    &\quad\quad + \Delta n_{1y}(\boldsymbol{\Omega}_a)n_{2y}(\boldsymbol{\Omega}_b)\big] \nonumber \\
    &\quad - U_1 \big[n_{1x}(\boldsymbol{\Omega}_a)n_{1x}(\boldsymbol{\Omega}_b)
    + n_{1z}(\boldsymbol{\Omega}_a)n_{1z}(\boldsymbol{\Omega}_b) \nonumber \\
    &\quad\quad + n_{2x}(\boldsymbol{\Omega}_a)n_{2x}(\boldsymbol{\Omega}_b)
    + n_{2z}(\boldsymbol{\Omega}_a)n_{2z}(\boldsymbol{\Omega}_b)\big] \bigg\} \nonumber \\
    &\quad - h\int d\boldsymbol{\Omega}\,
    \hat{\boldsymbol{\Psi}}^\dagger\tau^x\sigma^0\hat{\boldsymbol{\Psi}} .
\end{align}
The Haldane pseudopotentials are $V_0=1$ for $U_0$, $V_0=0.713202$ and $V_1=0.016740$ for $U_1$, and $V_0=0.571813$ and $V_1=0.064084$ for $U_2$.
The remaining parameters are $\Delta=0.011030$ and $h=0.147614$.
The conversion from the Haldane pseudopotentials $V_0,V_1$ to the spherical-harmonic coefficients $U_l^{0,1,2}$ of the interaction potential $U_{0,1,2}(\boldsymbol{\Omega}_a,\boldsymbol{\Omega}_b)$ follows Eq.~(A13) of Ref.~\cite{Han24}.

\begin{figure}[h]
\includegraphics[width=\columnwidth]{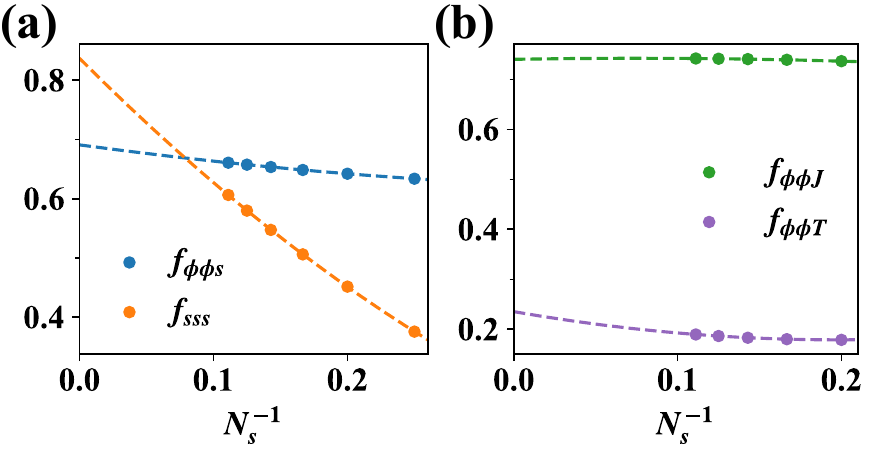}
\caption{Finite-size extrapolation of bulk $O(2)$ OPE coefficients. (a) $f_{\phi\phi s}$ and $f_{sss}$. (b) $f_{\phi\phi T}$ and $f_{\phi\phi J}$.}
\label{fig:app_bulk_o2_ope}
\end{figure}

As a check of the bulk realization, Fig.~\ref{fig:app_bulk_o2_ope} shows the finite-size extrapolation of four OPE coefficients.
The extrapolated values are listed in Table~\ref{tab:app_bulk_o2_ope} and agree well with the conformal bootstrap results.
The reported uncertainties are estimated by comparing fits with and without the largest included system size.
For $f_{\phi\phi J}$ and $f_{\phi\phi T}$, we use Eq.~(A32) of Ref.~\cite{Han24}, together with the values of $\Delta_\phi$, $C_J/C_J^{\rm free}$, and $C_T/C_T^{\rm free}$ from Ref.~\cite{Chester2020CarvingExponents}, to obtain the corresponding entries in Table~\ref{tab:app_bulk_o2_ope}.

\begin{table}[h]
\caption{Bulk $O(2)$ OPE coefficients.}
\label{tab:app_bulk_o2_ope}
\begin{ruledtabular}
\begin{tabular}{lcccc}
          & $f_{\phi\phi s}$ & $f_{sss}$ & $f_{\phi\phi J}$ & $f_{\phi\phi T}$ \\
\hline
This work & 0.6909(7) & 0.8371(36) & 0.7403(10) & 0.2352(8) \\
Bootstrap~\cite{Chester2020CarvingExponents} & 0.6871 & 0.8309 & 0.7435 & 0.2313 \\
\end{tabular}
\end{ruledtabular}
\end{table}

\section{Additional conformal multiplets}
\label{app:extra_multiplets}

\begin{figure}[h]
\includegraphics[width=\columnwidth]{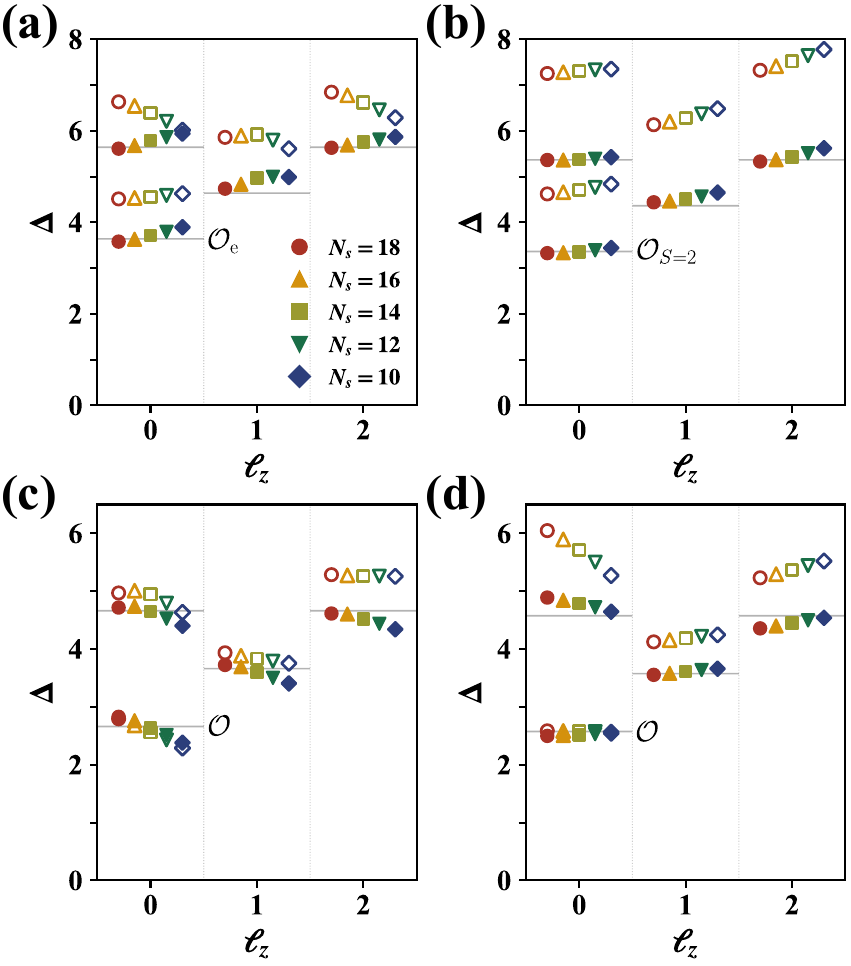}
\caption{Additional conformal multiplets for surface primaries discussed in the main text. Open symbols show raw dimensions calibrated by the protected displacement operator, $\Delta_D=3$. Solid symbols include the leading correction from the irrelevant boundary perturbation associated with $D$. Panels (a) and (b) supplement the normal surface spectra, with (a) the lowest $\mathbb{Z}_2$-even primary above $D$ in the $O(2)$ normal surface CFT and (b) the lowest $S=2$ primary in the $O(3)$ normal surface CFT. Panels (c) and (d) supplement the ordinary surface spectra, with (c) the lowest $\mathbb{Z}_2$-odd $O(2)$-singlet primary and (d) the lowest $\mathbb{Z}_2$-even $O(3)$ traceless-tensor primary.}
\label{fig:5}
\end{figure}

Fig.~\ref{fig:5} shows four additional multiplets that are used in the summary tables but are not displayed in the main figures.
The purpose is to support the primary assignments by their descendant structure.
A boundary primary $\mathcal{O}$ with dimension $\Delta_{\mathcal{O}}$ generates states
\begin{equation}
    \partial^m\bar\partial^{\bar m}\mathcal{O},
    \qquad
    \Delta=\Delta_{\mathcal{O}}+m+\bar m,
    \qquad
    l_z=m-\bar m .
\label{eq:app_descendant_pattern}
\end{equation}
Thus the identification uses a whole conformal multiplet rather than a single state.
At finite $N_s$, the raw levels drift away from exact integer spacing, which are shown by hollow symbols.
As in the main text, we remove the leading finite-size correction from the irrelevant boundary perturbation associated with $D$, using the procedure in Appendix~\ref{app:CPT}.
The corrected levels in Fig.~\ref{fig:5} show the expected integer-spaced pattern within the remaining finite-size drift.

For the normal boundary condition, Fig.~\ref{fig:5}(a) gives the second unprotected primary (lowest $\mathbb{Z}_2$-even primary above $D$) in the $O(2)$ normal surface CFT, denoted $\mathcal{O}_{\rm e}$, with
\begin{equation}
    \Delta_{\mathcal{O}_{\rm e}}=3.58(5).
\end{equation}
Fig.~\ref{fig:5}(b) gives the lowest $S=2$ primary in the $O(3)$ normal surface CFT, denoted $\mathcal{O}_{S=2}$, with
\begin{equation}
    \Delta_{\mathcal{O}_{S=2}}=3.33(11).
\end{equation}
Together with the conformal multiplets for normal surface CFTs shown in Fig.~\ref{fig:Fig1} and Fig.~\ref{fig:Fig2}, these results show that the observed unprotected primaries lie above the displacement operator.
This supports the statement in the main text that, in the low-lying normal spectrum, the only primary below $D$ is the protected tilt operator $t$.

For the ordinary boundary condition, Fig.~\ref{fig:5}(c) gives the lowest $\mathbb{Z}_2$-odd $O(2)$-singlet primary, with
\begin{equation}
    \Delta_{\mathcal{O}}=2.83(7),
\end{equation}
and Fig.~\ref{fig:5}(d) gives the lowest $\mathbb{Z}_2$-even $O(3)$ traceless-tensor primary, with
\begin{equation}
    \Delta_{\mathcal{O}}=2.499(8).
\end{equation}
Both dimensions are above the boundary relevance threshold $\Delta=2$.
They therefore do not change the ordinary fixed point, whose only relevant surface operator is the vector field $\hat\phi$ discussed in the main text.
The main value of these extra towers is that they resolve additional symmetry sectors of the ordinary surface CFT.

\section{Real-space boundary cut}
\label{app:real_space_cut}

In the thermodynamic limit, the orbital-space boundary in Eq.~\eqref{eq:surface_orbital} should approach a real-space cut at the equator.
At finite $N_s$, however, each lowest-Landau-level orbital has a finite angular width, so the orbital-space cut is not perfectly local.
We therefore check the construction using a boundary field that is local in real space.
The corresponding surface Hamiltonian is
\begin{equation}
\label{eq:real_space_surface}
    H_{\rm surface}^{\rm real}=
    \sum_{\tau=1}^{2}\int_{\theta>\pi/2} d\Omega\,
    \boldsymbol{h}_{\tau}^{s}\cdot\boldsymbol{n}_{\tau}(\Omega) .
\end{equation}
For the normal boundary condition we choose
$\hat{\boldsymbol{h}}_1^s=-\hat{\boldsymbol{h}}_2^s=\hat{z}$.

\begin{figure}[h]
\includegraphics[width=\columnwidth]{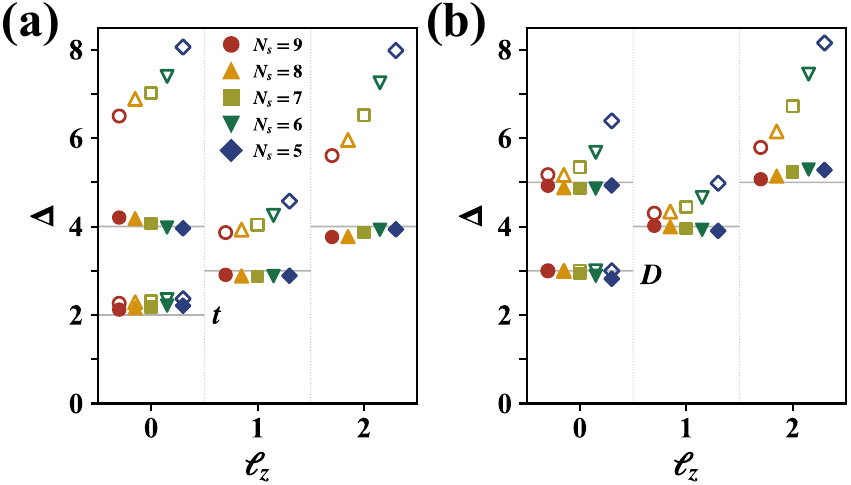}
\caption{Conformal multiplets of (a) $t$ and (b) $D$ in the $O(3)$ normal surface CFT obtained from the real-space boundary cut.}
\label{fig:6}
\end{figure}

Fig.~\ref{fig:6} shows the multiplets of $t$ and $D$ for the $O(3)$ normal surface CFT, with states included up to $\Delta=5$ and $l_z=2$.
The accessible system sizes are smaller than those in the orbital-space calculation, and the raw spectra therefore show larger finite-size drifts than in Fig.~\ref{fig:Fig2}(a) and Fig.~\ref{fig:Fig2}(b).
After the same leading correction from $D$ is included, the integer spacing of the conformal multiplets is recovered.
This supports the conclusion that the real-space and orbital-space boundary cuts flow to the same surface universality class.
The orbital-space cut is more efficient numerically because the $m<0$ orbitals can be integrated out, which gives access to larger $N_s$.

\section{Calibration using bulk stress tensor}
\label{app:calibration}

\begin{figure}[h]
\includegraphics[width=\columnwidth]{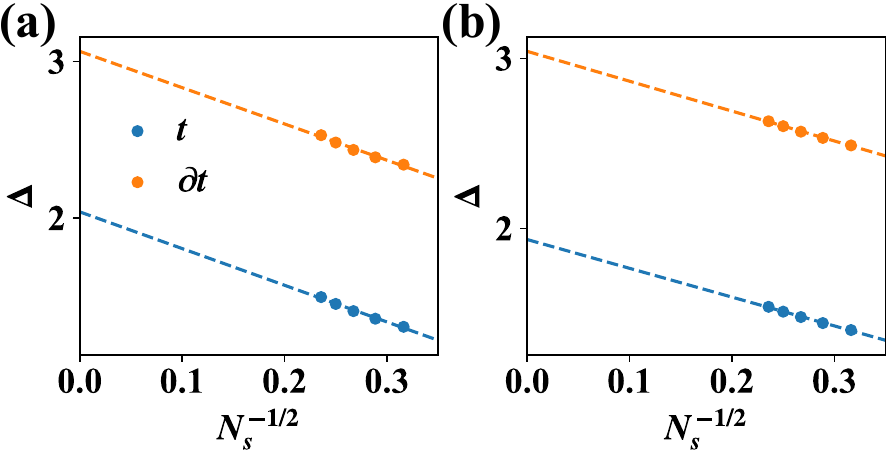}
\caption{The scaling dimension of $t$ and $\partial t$ in (a) $O(2)$ and (b) $O(3)$ normal surface CFTs, calibrated using the bulk stress tensor $\Delta_{T_{\mu \nu}}=3$.}
\label{fig:7}
\end{figure}

As stated in \cite{Note1}, calibrating the surface operator dimensions using the bulk stress tensor $\Delta_{T_{\mu \nu}}=3$ should yield consistent result, because the speed of light should be the same for the bulk CFT and the corresponding surface CFT~\cite{hu24,Zhou25}.
In Fig.~\ref{fig:7}, we have computed the scaling dimensions using the bulk speed of the light.
According to Ref.~\cite{Zhou25}, the calculated scaling dimensions in finite-size systems are subject to corrections from irrelevant surface and bulk primaries in the symmetry-singlet sector
\begin{equation}
    \Delta_{\hat{\phi}}(N) = \Delta_{\hat{\phi}} + \sum_{\hat{S}} \lambda_{\hat{\phi}\hat{S}} R^{2-\Delta_{\hat{S}}} + \sum_{S} \lambda_{\hat{\phi}S} R^{3-\Delta_S} .
\end{equation}
In Fig.~\ref{fig:7} we only did linear fitting in $N^{-1/2}$.
This amounts to only considering the perturbation from irrelevant surface primary $D$.
One can see that this recovers the expected integer-spacing spectra: for $O(2)$ the result is $\Delta_t=2.04(4)$ and $\Delta_{\partial t}=3.07(6)$; for $O(3)$ the result is $\Delta_t=1.93(2)$ and $\Delta_{\partial t}=3.04(2)$.
The scaling dimensions for $t$ and $\partial t$ agree well with expectation $\Delta_t=2$ and $\Delta_{\partial t}=3$.

\section{Conformal perturbation theory}
\label{app:CPT}

The finite-size spectra receive corrections from irrelevant surface and bulk primaries in the symmetry-singlet sector.
The bulk pseudopotentials are chosen to be close to the bulk fixed point.
We therefore assume that the leading bulk irrelevant drift is subdominant and model the dominant drift by the leading irrelevant surface perturbation, the displacement operator $D$ with $\Delta_D=3$.
This is the same conformal-perturbation strategy used in Ref.~\cite{Zhou25}.
For completeness, we give the cost function used in the present work:
\begin{equation}
    \begin{aligned}
\label{eq:cost}
    \mathcal{L} &= \sum_{\mathcal{O}}\sum_{\substack{0\le\bar{m}\le m\le3 \\ m+\bar{m}\le3}}
    \bigg[\frac{E^{(\rm num)}(\partial^m\bar{\partial}^{\bar{m}}\mathcal{O})}{E_0} \\
    &\quad
    -\delta E(\partial^m\bar{\partial}^{\bar{m}}\mathcal{O},\lambda_{\mathcal{O}D},\Delta_{\mathcal{O}}^{(\rm ref)},3) - (\Delta_{\mathcal{O}}+m+\bar{m})\bigg]^2  \\
    &\quad + \sum_{\substack{0\le\bar{m}\le m\le3 \\ m+\bar{m}\le3}}
    \bigg[\frac{E^{(\rm num)}(\partial^m\bar{\partial}^{\bar{m}}D)}{E_0}
    \\
    &\quad -\delta E(\partial^m\bar{\partial}^{\bar{m}}D,\lambda_{DD},3,3)  - (3+m+\bar{m})\bigg]^2 .
\end{aligned}
\end{equation}
Here $E_0$ converts the numerical energy scale to scaling dimensions, and $\delta E$ is the first-order shift induced by the $D$ perturbation.
Readers can refer to Eq.(B.9) in Ref.~\cite{Zhou25} for the detailed form of $\delta E$.
For each $N_s$, we minimize Eq.~\eqref{eq:cost} with respect to $E_0$, the scaling dimensions $\Delta_{\mathcal{O}}$, and the couplings $\lambda_{\mathcal{O}D}$ and $\lambda_{DD}$.
The index $\mathcal{O}$ runs over all fitted primaries except $D$.

The corrected spectra shown in Fig.~\ref{fig:Fig1}, Fig.~\ref{fig:Fig2}, Fig.~\ref{fig:Fig3}, Fig.~\ref{fig:5}, and Fig.~\ref{fig:6} are consistent with the expected integer spacing of boundary descendants.
This provides an internal check that the leading finite-size drift is captured by the $D$ perturbation.
As a side remark, we emphasize that while $\Delta_D=3$ was forced in the reported raw data (hollow symbols in the reported figures for all conformal multiplets), it is not ensured within the conformal perturbation scheme here.
At the same time, Eq.~\eqref{eq:cost} does not include all possible subleading irrelevant operators, as well as possible perturbation from irrelevant bulk primaries.
The quoted uncertainties in the main text should therefore be viewed as finite-size estimates rather than purely statistical error bars.
A more complete conformal-perturbation analysis, including additional irrelevant primaries, would be a useful improvement for future work.

\end{document}